\documentclass[10pt]{article}
\usepackage{fullpage}
\usepackage{graphics}
\usepackage{graphicx}
\usepackage{epsfig}
\usepackage{dcolumn}
\usepackage{bm}
\usepackage{amssymb}
\usepackage{amsmath}
\usepackage{amsfonts}
\usepackage{color}

\def\red{\textcolor{black}}

\def\##1{\underline{#1}}
\def\##1{{\bf{#1}}}
\def\=#1{\underline{\underline{#1}}}

\def\+#1{\underline{\bf #1}}
\def\*#1{\underline{\underline{\bf #1}}}

\def\r#1{(\ref{#1})}
\def\l#1{\label{#1}}
\def\c#1{\cite{#1}}

\def\le{\left(}
\def\ri{\right)}
\def\les{\left[}
\def\ris{\right]}
\def\lec{\left\{}
\def\ric{\right\}}

\def\.{\mbox{ \tiny{$^\bullet$} }}

\def\eps{\epsilon}

\def\epso{\epsilon_{\scriptscriptstyle 0}}

\def\muo{\mu_{\scriptscriptstyle 0}}

\def\ko{k_{\scriptscriptstyle 0}}

\begin{document}
\begin{center}

\Large{ Dynamically controllable, \red{homogeneous}, anisotropic metamaterials with simultaneous attenuation and amplification
}
\end{center}

\begin{center}
\vspace{10mm} \large

 Tom G. Mackay\footnote{E--mail: T.Mackay@ed.ac.uk.}\\
{\em School of Mathematics and
   Maxwell Institute for Mathematical Sciences\\
University of Edinburgh, Edinburgh EH9 3FD, UK}\\
and\\
 {\em NanoMM~---~Nanoengineered Metamaterials Group\\ Department of Engineering Science and Mechanics\\
Pennsylvania State University, University Park, PA 16802--6812,
USA}\\
 \vspace{3mm}
 Akhlesh  Lakhtakia\footnote{E--mail: akhlesh@psu.edu}\\
 {\em NanoMM~---~Nanoengineered Metamaterials Group\\ Department of Engineering Science and Mechanics\\
Pennsylvania State University, University Park, PA 16802--6812, USA}

\normalsize

\end{center}

\vspace{10mm}

\begin{abstract}
Anisotropic homogeneous metamaterials that are neither wholly dissipative nor wholly active at a specific frequency are permitted by classical electromagnetic theory.  Well-established homogenization formalisms indicate that such a metamaterial may be realized quite simply as a random mixture of electrically small (possibly nanoscale) spheroidal particles  of  at least two different isotropic dielectric materials, one of which must be dissipative but the other active.   The dielectric properties of this metamaterial are influenced by the volume fraction,  spatial distribution, particle shape and size, and the relative permittivities of the component materials. Similar metamaterials with more complicated linear as well as nonlinear constitutive properties  are possible. Dynamic control of the active component material,  for example via stimulated Raman scattering, affords dynamic control of the metamaterial.
 \end{abstract}



\section{Introduction}

Causality mandates that electromagnetic fields must attenuate as they propagate inside homogeneous,
passive  linear materials \c{Zangwill}. Weak electromagnetic fields in certain spectral regimes  can be amplified in some homogeneous linear materials,
provided that strong electromagnetic fields can pump in the energy needed for amplification
 \cite{CS2,Bespalov,Silicon1,Silicon2}.
Conditions on the constitutive parameters of homogeneous linear materials have been derived to determine if a linear material is either dissipative  or  active, but not both, at a specific frequency \c{Kong-Book}.

Active component materials feature prominently in the field of metamaterials \c{Hess}, in order to overcome losses \c{Dong_APL} and to enhance performance \c{Strangi}.
While anisotropic structures containing active components have been reported upon previously \c{Tretyakov,Sun_APL,Savelev},  the prospect of
 simultaneous   attenuation and amplification of electromagnetic fields in  homogeneous materials at a specific frequency, depending upon the orientation of electric fields, has not been considered hitherto.
Materials supporting amplification and attenuation at the same frequency are necessarily anisotropic. A host of  applications for  them can be envisaged. These materials could be used in directional coupling devices \c{Directional_coupling}  as well as in spatially discriminatory  and/or  frequency--discriminatory optical amplifiers \c{Optical_amp}.  Amplification could be dynamically controlled by exploiting, for instance,
the phenomenon of stimulated Raman scattering \cite{CS2,LR2008}, thereby affording dynamic control.
That is, if the active component material were Raman active, then the degree of amplification achieved for a probe laser beam at a desired frequency could be dynamically controlled  by means of a strong pump laser beam at  a determined frequency which induces Raman transitions. The difference between the probe and pump laser frequencies is specified  by the energy levels of the Raman-active material.
 The ability to
 suppress radiation leakage in certain directions while promoting propagation in other directions could be harnessed to amplify
surface--plasmon polaritons and reduce optical noise in biosensing applications \c{Berini}, for example. Radomes for enhancing or reducing directionality of radiation from optical antennas could be made of these materials, leading to
enhancements in the
efficiency of photodetection, light emission, and sensing \c{Novotny}.

Motivated by these potential applications, here we propose dynamically controllable anisotropic materials   which simultaneously exhibit both dissipation and amplification at a specific frequency, depending upon the orientation of electric fields.

 \section{Simultaneous attenuation and amplification}

Consider a generally anisotropic, homogeneous, linear, dielectric material  characterized by the frequency-domain
 constitutive relations
 \begin{equation}
 \left.
 \begin{array}{l}
 \#D (\#r,\omega) = \epso\, \=\eps(\omega) \. \#E (\#r,\omega) \vspace{4pt} \\
  \#B (\#r,\omega) = \muo \#H (\#r,\omega) \end{array} \right\},
  \end{equation}
 where $\=\eps(\omega)$ is the relative permittivity dyadic \cite{EAB,HC_Chen}
 at angular frequency $\omega$, and $\epso$ and $\muo$ are, respectively, the permittivity and permeability of free space.  The time-averaged dissipated power per unit volume is given by \cite{HC_Chen}
 \begin{equation} \l{gQ}
 Q (\#r,\omega)  = -\frac{i \omega \epso}{4} \, \#E^*(\#r,\omega) \. \les \=\eps(\omega) - \={\tilde{\eps}}(\omega) \ris \. \#E (\#r,\omega),
  \end{equation}
  where  $i=\sqrt{-1}$; an $\exp(-i\omega t)$ dependence on time $t$ is implicit; the superscript ${}^*$ denotes the complex conjugate; and
  $\={\tilde{\eps}}(\omega)$ is the hermitian conjugate of $\=\eps(\omega)$.

At a specific angular frequency $\omega$,
the chosen material is classified as \c{Tan_MOTL}:
   \begin{itemize}
\item   dissipative if $ Q (\#r,\omega)  > 0$, which  requires
 the dyadic $i\les \=\eps(\omega) - \={\tilde{\eps}}(\omega) \ris$ to be negative definite; or
\item   active if $ Q(\#r,\omega) <0$, which   requires
 the dyadic $i\les \=\eps(\omega) - \={\tilde{\eps}}(\omega) \ris$ to be positive definite.
 \end{itemize}
 All  eigenvalues of a negative/positive definite dyadic are negative/positive \cite{Lutkepohl}.  Henceforth,
 for compact representation,
 the dependency of $Q$ and   $\#E $ on $\omega$ and $\#r$,  will not be explicitly stated; similarly,
 the dependency of $\=\eps$  (and its components)   on $\omega$   will not be explicitly stated.

The foregoing classification   fails to accommodate to the prospect of  $i\le \=\eps  - \={\tilde{\eps}}  \ri$ being  indefinite \cite{Lutkepohl}---i.e., when some but not all eigenvalues of $i\le \=\eps - \={\tilde{\eps}}  \ri$ are positive, the remaining eigenvalues being negative. For orthorhombic materials
\c{Nye},
$i\le\=\eps  - \={\tilde{\eps}}  \ri$ is indefinite provided that
 $\mbox{Im} \lec \=\eps   \ric$ is  indefinite, where the operator $\mbox{Im} \lec \cdot \ric$ delivers the imaginary part.

 The simplest material, at least from a mathematical perspective, for which $\mbox{Im} \lec \=\eps \ric$ is  indefinite
 is a uniaxial dielectric material
 whose relative permittivity dyadic has the form \c{HC_Chen}
 \begin{equation} \l{uniaxial}
 \=\eps = \eps^\perp \le \=I - \hat{\#u}  \hat{\#u} \ri
+ \eps^\parallel \hat{\#u}  \hat{\#u},
 \end{equation}
 with $
 \mbox{Im} \lec \eps^\perp \ric \,  \mbox{Im} \lec \eps^\parallel \ric < 0$.
 Herein $\=I$ is the identity   dyadic and the unit vector $\hat{\#u}$ is parallel to the material's optic axis.
 For this  material,
 it may be inferred from Eq.~\r{gQ} that
 $Q   $ for $\#E$ directed along $\hat{\#u}$ has the opposite sign to
 $Q$ for $\#E$ directed perpendicular to $\hat{\#u}$, because
  \begin{eqnarray}
\nonumber
 Q&=&\frac{ \omega \epso}{2} \Big[
  \mbox{Im} \lec \eps^\perp \ric\#E^*\. \le \=I - \hat{\#u}  \hat{\#u} \ri\.\#E
  \\&&\qquad+
  \mbox{Im} \lec \eps^\parallel\ric\#E^*\.   \hat{\#u}  \hat{\#u} \.\#E
\Big]\,.
   \label{eQ1}
\end{eqnarray}
Thus,
  there is dissipation associated with  certain orientations of the electric field but
  amplification  with other orientations.

\section{Realization as a homogenized composite material}

Is it possible to realize a uniaxial dielectric material  for which $\mbox{Im} \lec \=\eps \ric$ is  indefinite?
We now demonstrate, using well--established theoretical formalisms based on the homogenization of particulate composite materials, that  materials with indefinite $\mbox{Im} \lec \=\eps \ric$ may be conceptualized as homogenized composite materials (HCMs). As these engineered
 materials will  simultaneously exhibit  both amplification and dissipation at the same frequency, they should more properly be called as  metamaterials \cite{Walser}. In the visible spectrum (380--770 THz), the maximum linear dimensions of the  component particles are required to be less than $\sim$30~nm \c{vdH}.

Consider a composite material that is a  mixture of two component materials
labeled  `a' and `b'. Dispersed randomly as
 identically oriented, electrically small, conformal, spheroids, both component materials are isotropic dielectric materials
 with relative permittivities $\eps_a$ and $\eps_b$.
 The surface of a  spheroid, relative to its centroid, is prescribed by the position vector
$\#r = \rho \, \=U \. \hat{\#r}_{s}$,
where  $\rho>0$ is a linear measure of particle size, the shape dyadic
\begin{equation}
\=U = \frac{1}{\sqrt{U}} \le \=I - \hat{\#u}   \hat{\#u} \ri
+ U \hat{\#u} \hat{\#u}, \qquad \le U  >0 \ri
\end{equation}
contains the shape parameter $U$,
and the unit vector $\hat{\#r}_{s} $ prescribes the surface of the concentric unit sphere.
Oblate spheroidal particles are characterized by  $U\in(0,1)$, prolate
spheroidal particles by $ U > 1$, and spherical
particles by $U=1$.
 The volume fraction of component material `a' is $f_a$, while that of component material `b' is $f_b = 1 - f_a$.
 The spheroidal  shape and the identical alignment of the component particles endows the  HCM with uniaxial symmetry
 \cite{EAB,HC_Chen}; i.e., the relative
 permittivity dyadic of the HCM  has the form
 \begin{equation} \l{uniaxial_HCM}
 \=\eps_{HCM} = \eps^\perp_{HCM} \le \=I - \hat{\#u} \, \hat{\#u} \ri
+ \eps^\parallel_{HCM} \hat{\#u} \, \hat{\#u}.
 \end{equation}
Porous columnar-thin-film
sections and nanoparticle arrays can be fabricated to realize such HCMs \c{Hodgkinson_book,Mondal}.

\section{Estimates of constitutive parameters}

Let us now present representative numerical estimates of $ \=\eps_{HCM}$, as yielded by the Bruggeman formalism \c{WLM_MOTL}, the strong--permittivity--fluctuation theory (SPFT) \c{Kong,Z94}, and the Maxwell Garnett formalism \c{David,JW}; these estimates are identified by replacing the subscripts `HCM'  by `Br', `SPFT', and `MG', respectively.

\begin{figure}[!ht]
\centering
\includegraphics[width=7.0cm]{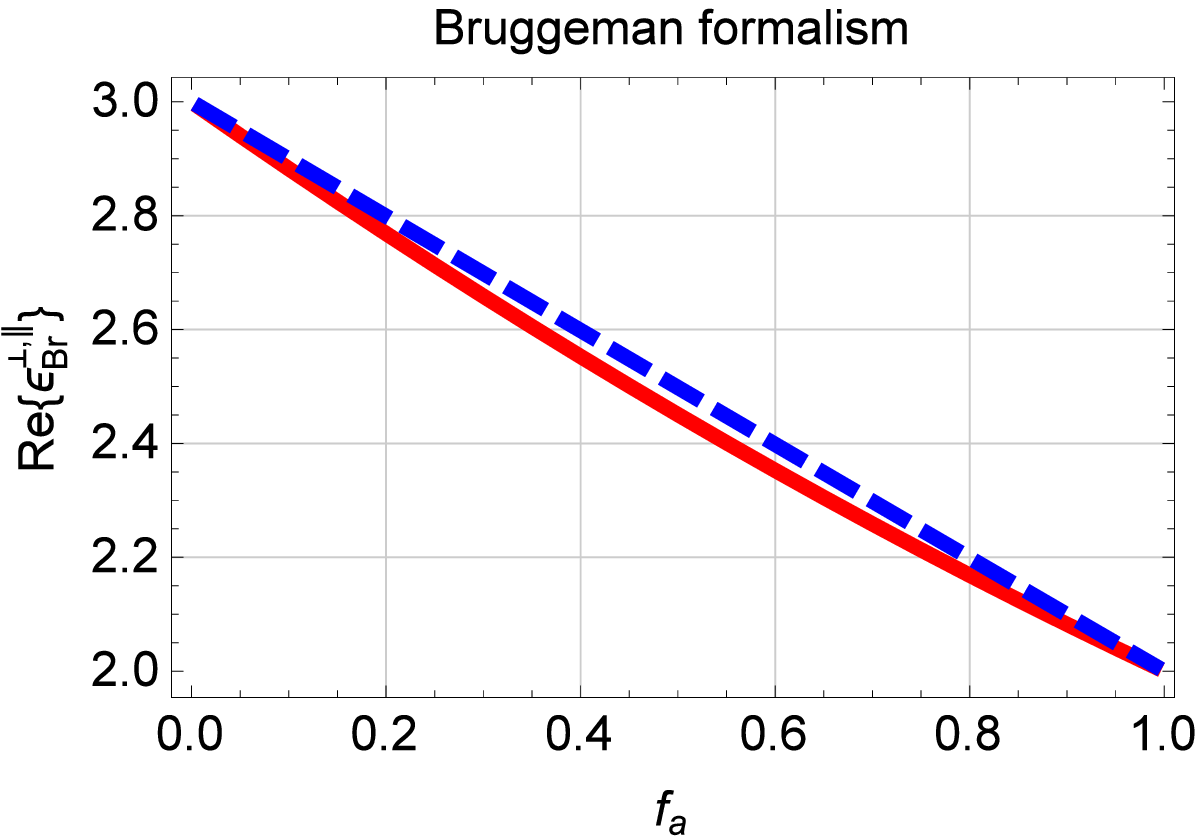}\\
\includegraphics[width=7.0cm]{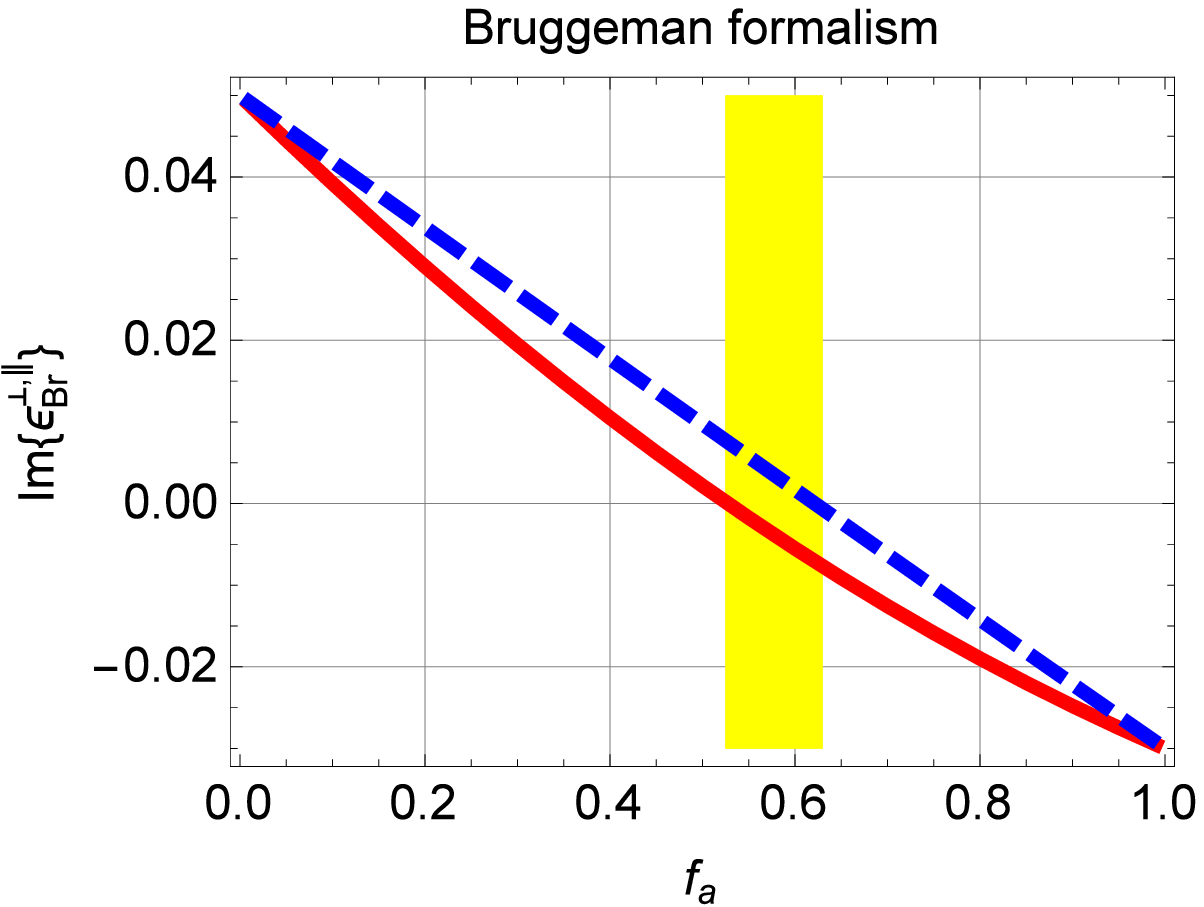}
 \caption{\label{eps_v_fa} (Color
  online) Real and
imaginary parts of $\eps^\perp_{Br}$ (red, solid curve) and
 $\eps^\parallel_{Br}$ (blue, dashed curve)
 plotted against the volume fraction $f_a$, when
 $\eps_a = 2 - 0.03 i $, $\eps_b = 3 + 0.05i $, and $U = 5$.
  The  yellow rectangle indicates the $f_a$-range where
 $\mbox{Im} \lec \eps^\perp_{Br} \ric < 0 $ but $\mbox{Im} \lec \eps^\parallel_{Br} \ric > 0 $.
  }
\end{figure}

For the purpose of illustration,
suppose that the component material `a' is an active material with $\eps_a   = 2 - 0.03 i $.
 This value of $\eps_a$ lies comfortably within the range typically employed for active components of metamaterials in the visible regime. For example,
a mixture of two commonly used  amplification
materials, namely Rhodamine 800  and  Rhodamine 6G,  possesses a relative permittivity with imaginary part in the range
$\le -0.15, -0.02 \ri$ and real part in the range $\le 1.8, 2.3 \ri$ across the frequency range 440--500 THz,
depending upon the relative concentrations  and the external pumping rate \c{Sun_APL}.
 Component material `b' is taken to be a dissipative material specified by $\eps_b = 3 + 0.05i $, and  the  shape parameter $U=5$.
The real and imaginary parts of the HCM's relative-permittivity scalars $\eps^\perp_{Br}$  and
 $\eps^\parallel_{Br}$, as estimated using the Bruggeman formalism \c{WLM_MOTL},
are plotted against the volume fraction $f_a$ in Fig.~\ref{eps_v_fa}. The real parts of $\eps^\perp_{Br}$  and
 $\eps^\parallel_{Br}$ decrease in an approximately linear manner from $3$ to $2$ as $f_a$ increases from $0$ to $1$.
Both $\mbox{Im} \lec \eps^{\perp}_{Br} \ric$ and $\mbox{Im} \lec \eps^{\parallel}_{Br} \ric$ decrease uniformly from $ \mbox{Im} \lec \eps_b \ric$ to $  \mbox{Im} \lec \eps_a \ric$ as $f_a$ increases from $0$ to $1$, but
 the decrease in $\mbox{Im} \lec \eps^{\perp}_{Br} \ric$ is more distinctly nonlinear than that in $\mbox{Im} \lec \eps^{\parallel}_{Br} \ric$.  The data show that
 \begin{itemize} \item[(i)] for  $0 < f_a < 0.52$, the HCM exhibits only dissipation, regardless of the orientation of the electric field  $\#E$; \item[(ii)] for $0.63 < f_a < 1 $, the HCM exhibits only amplification, regardless of the orientation of   $\#E$; and \item[(iii)]for  $0.52 < f_a < 0.63$, the HCM  exhibits simultaneously both  dissipation and amplification, depending on the orientation of  $\#E$.
\end{itemize}
Parenthetically, the HCM represented in Fig.~\ref{eps_v_fa} is not an `indefinite' material \c{EAB} since   $\mbox{Re} \lec \=\eps_{\,HCM} \ric$ is  positive definite.

\begin{figure}[!ht]
\centering
\includegraphics[width=7.0cm]{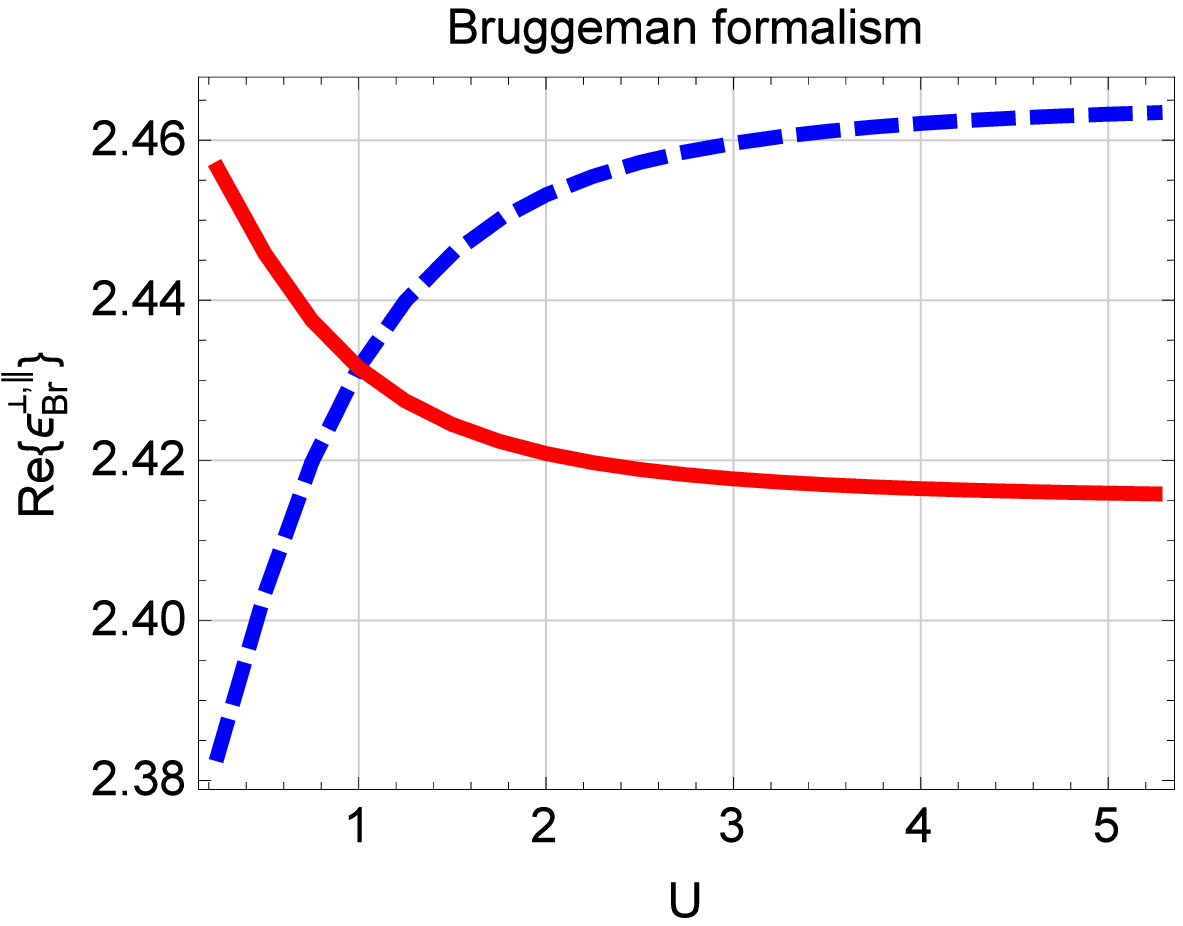}\\
\includegraphics[width=7.0cm]{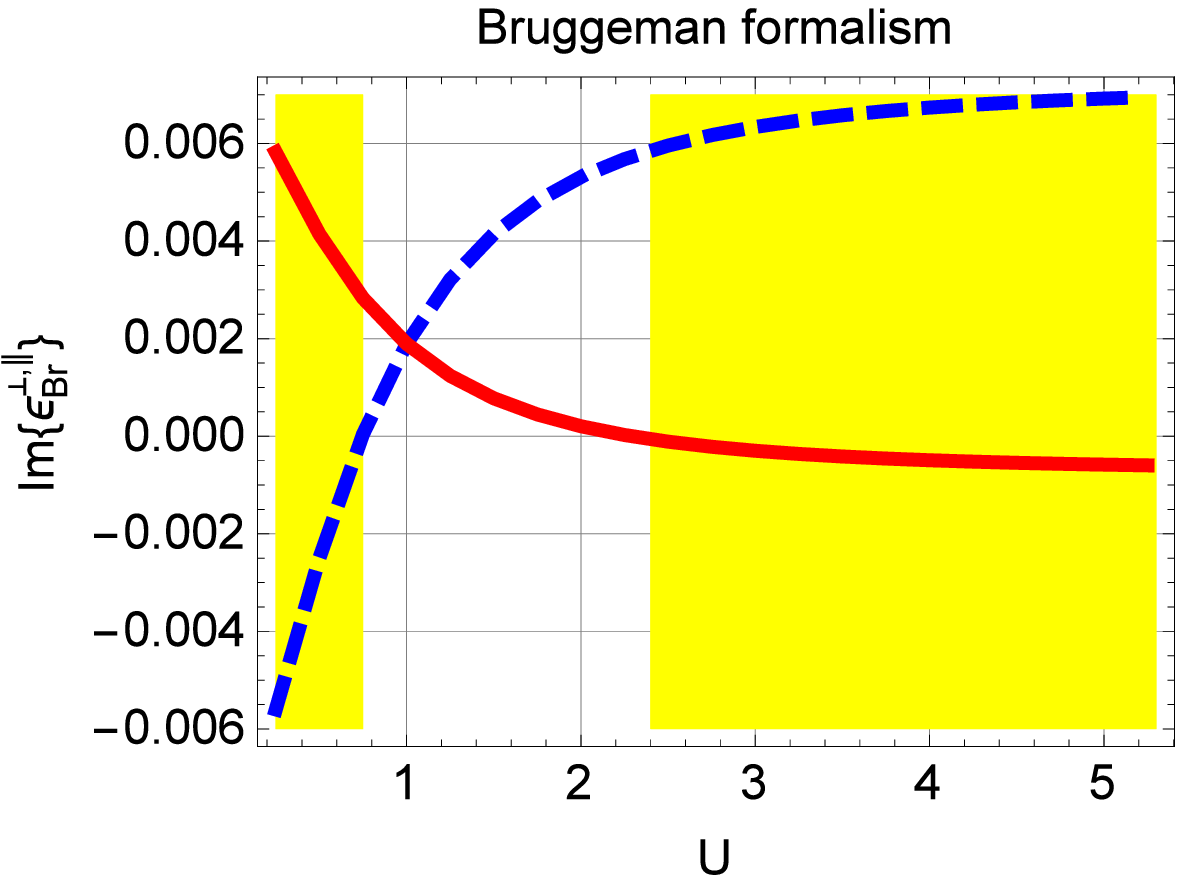}
 \caption{\label{eps_v_Ua}
(Color
  online) Real and
imaginary parts of $\eps^\perp_{Br}$ (red, solid curve) and
 $\eps^\parallel_{Br}$ (blue, dashed curve)
 plotted against the  shape parameter $U$, when
 $\eps_a = 2 - 0.03 i $, $\eps_b = 3 + 0.05i $, and $f_a = 0.535$.
  The two  yellow rectangles indicate the $U$-ranges where the product
 $\mbox{Im} \lec \eps^\perp_{Br} \ric \mbox{Im} \lec \eps^\parallel_{Br} \ric < 0 $.
  }
\end{figure}

The shape of the component particles plays an important role in the  simultaneous exhibition of dissipation and amplification. In Fig.~\ref{eps_v_Ua}, the real and imaginary parts of  $\eps^\perp_{Br}$  and
 $\eps^\parallel_{Br}$
 are plotted against   $U$, when $f_a = 0.535$ and the relative permittivities of the component materials are the same as for Fig.~\ref{eps_v_fa}.  The component particles are: oblate spheroids for $U < 1$, prolate spheroids for
$U > 1$, and  spheres for $U = 1$.
Both the real and imaginary parts of $\eps^{\perp}_{Br}$ decrease nonlinearly as $U$ increases whereas both the real and imaginary parts of $\eps^{\parallel}_{Br}$ increase nonlinearly as $U$ increases, with
 $\eps^{\perp}_{Br} = \eps^{\parallel}_{Br} $
 at $U=1$.  In the vicinity of $U=1$,   $\mbox{Im} \lec \eps^{\perp}_{Br} \ric$ and $\mbox{Im} \lec \eps^{\parallel}_{Br} \ric$ are both positive. However, for $U < 0.75$, we find that $\mbox{Im} \lec \eps^\parallel_{Br} \ric < 0 $ but $\mbox{Im} \lec \eps^\perp_{Br} \ric > 0 $.
Incidentally, the Bruggeman formalism for a particulate composite material yields the same results in the limit $U\to0$ as for a
periodically laminated composite material
\c{BW}.
Also, for $U > 2.4$, we find that $\mbox{Im} \lec \eps^\parallel_{Br} \ric > 0 $ but $\mbox{Im} \lec \eps^\perp_{Br} \ric < 0 $.
Thus, the HCM exhibits simultaneously both  dissipation and amplification provided that the component spheroidal particles are either sufficiently flattened or sufficiently elongated.

\begin{figure}[!ht]
\centering
\includegraphics[width=7.5cm]{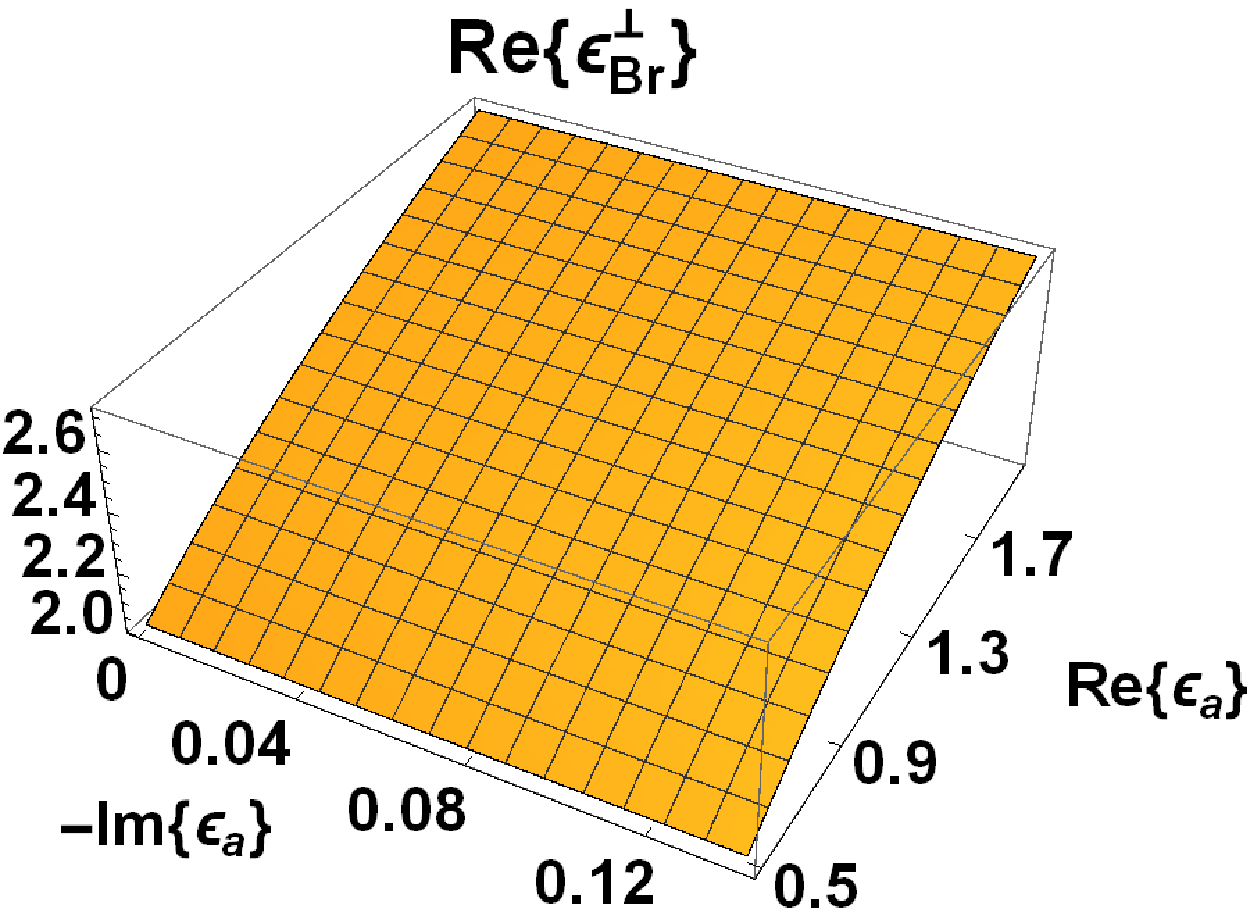}
\includegraphics[width=7.5cm]{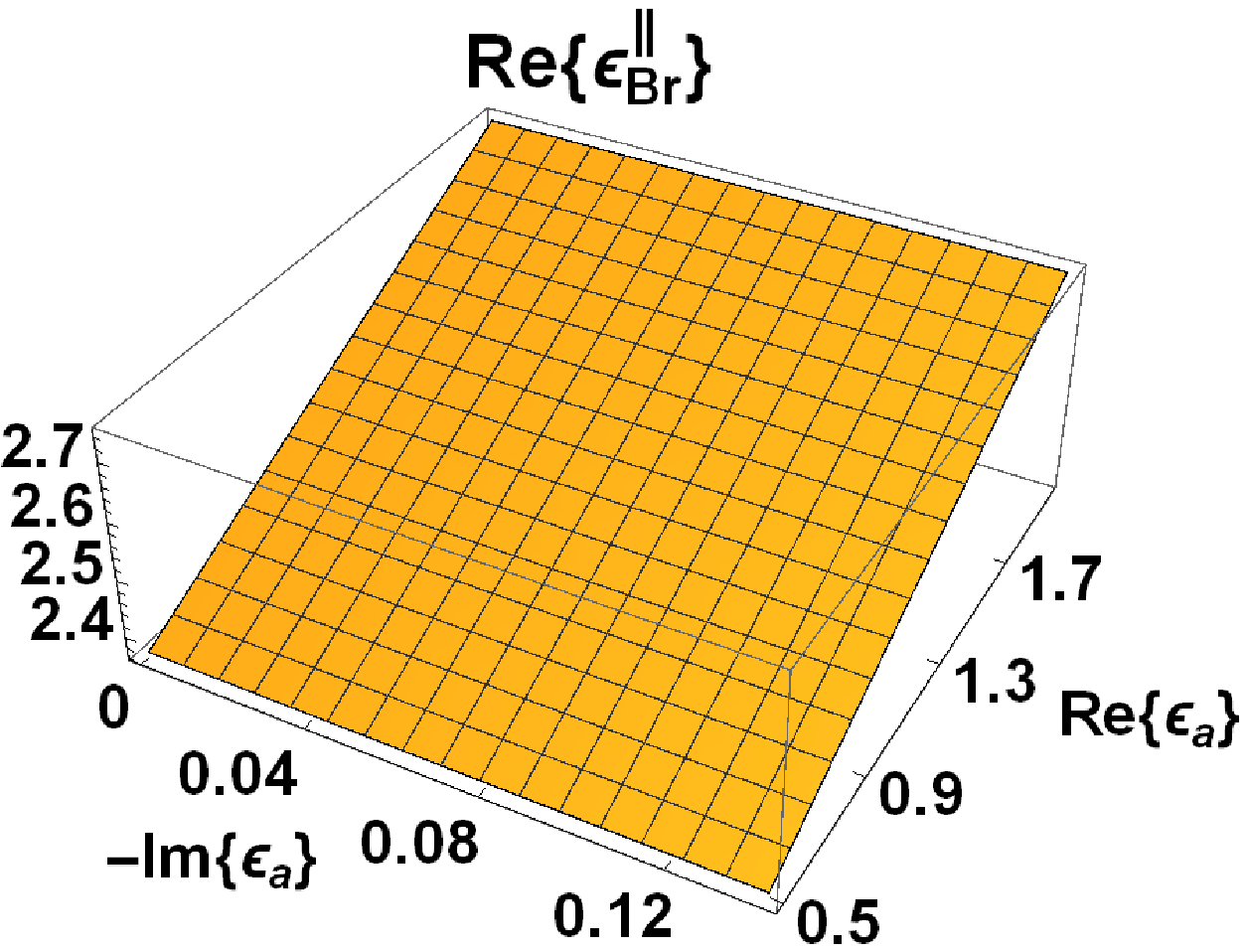}\\
\includegraphics[width=7.5cm]{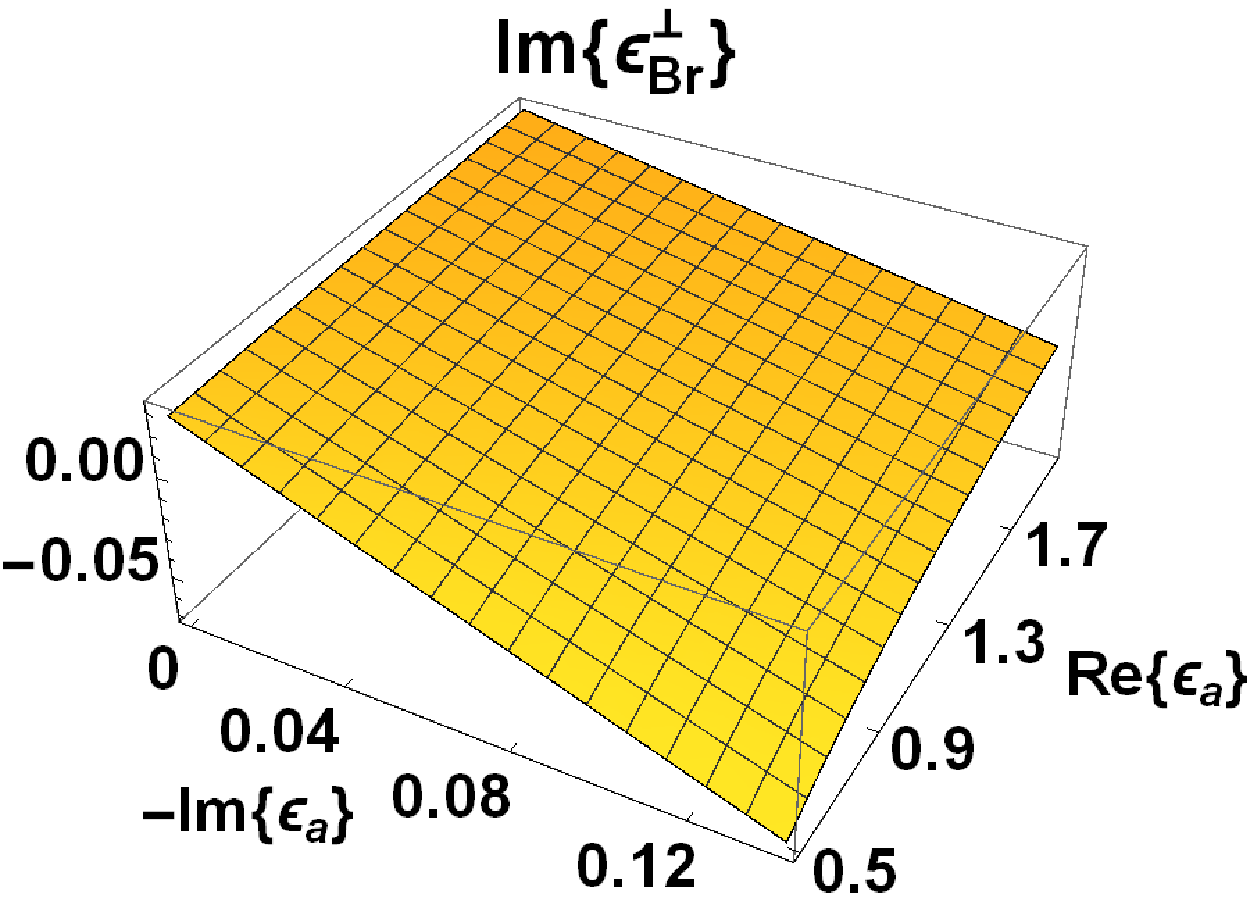}
\includegraphics[width=7.5cm]{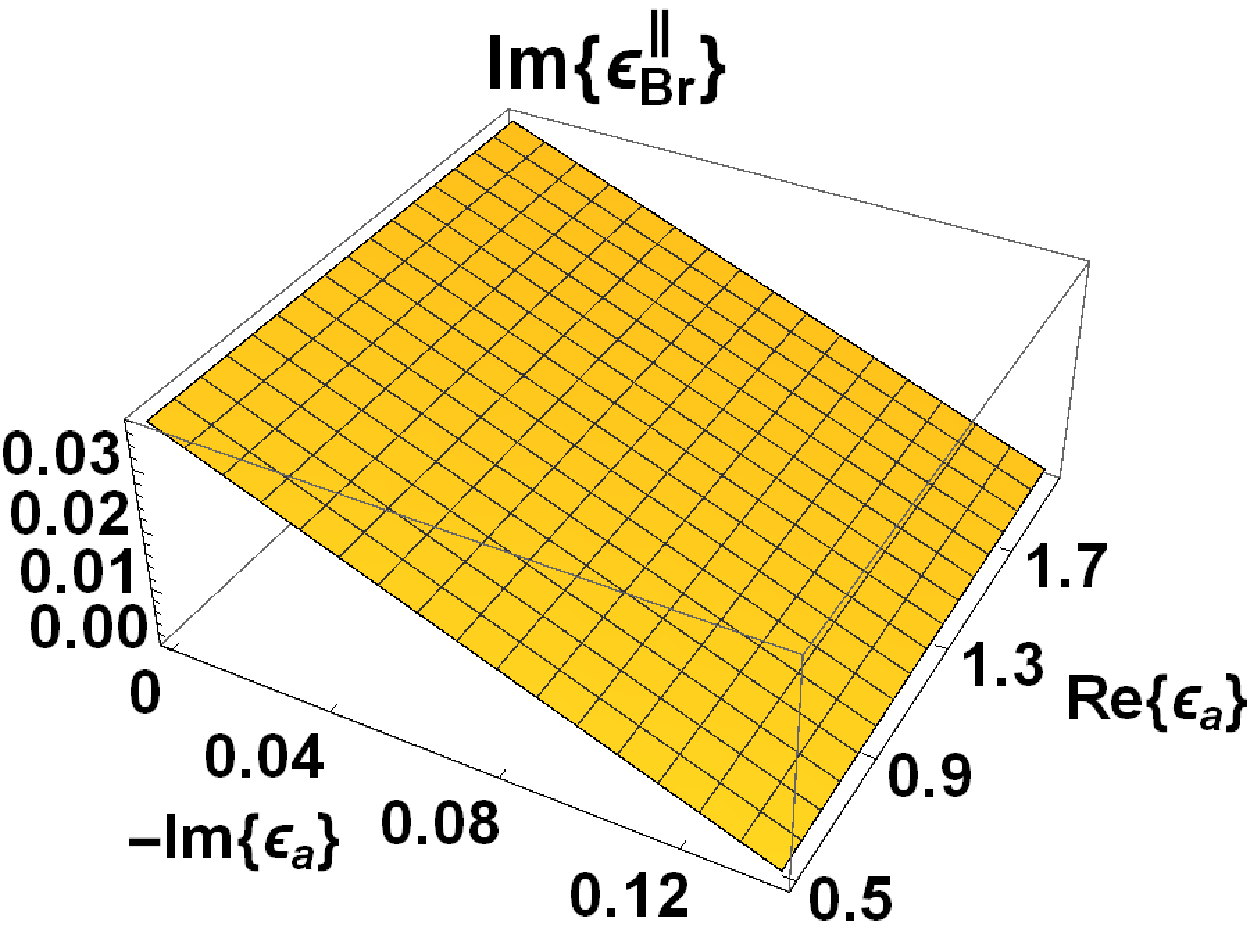}\\
\includegraphics[width=7.5cm]{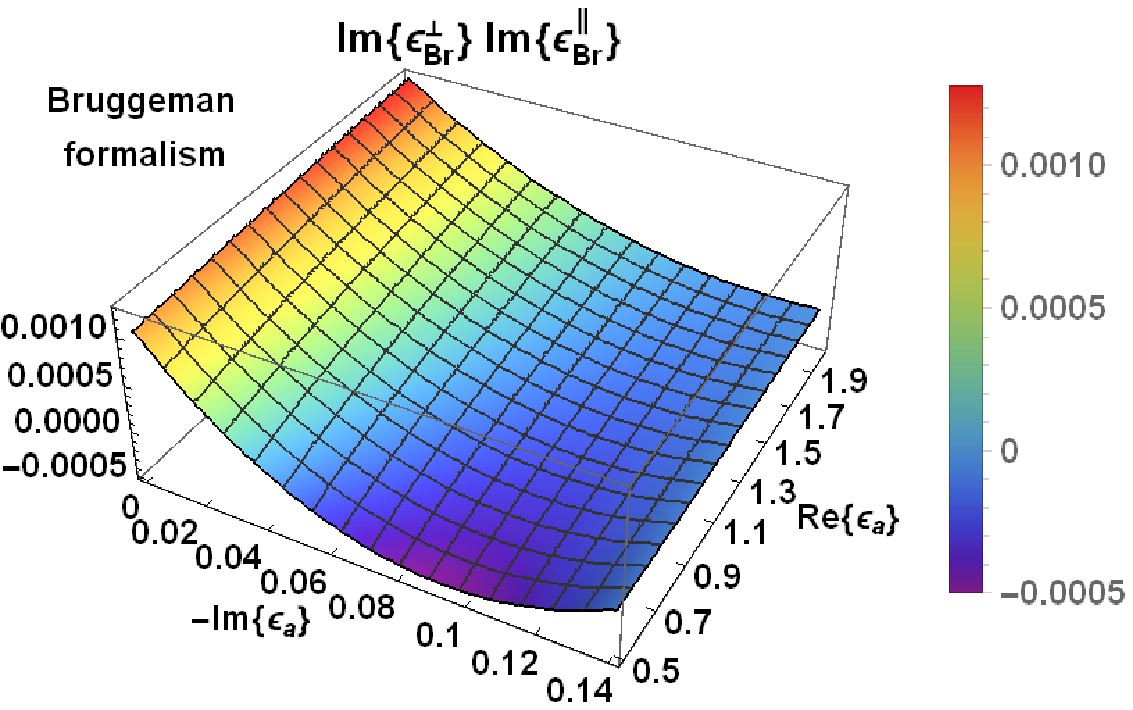}
 \caption{\label{eps_v_ea} (Color
  online)
  Real and imaginary parts $\eps^\perp_{Br}$  and
 $\eps^\parallel_{Br}$, as well as
 the product  $\mbox{Im} \lec \eps^\perp_{Br} \ric \mbox{Im} \lec \eps^\parallel_{Br} \ric $,
 plotted against
  the real and (negative) imaginary parts of the relative permittivity $\eps_a$, when
 $\eps_b = 3 + 0.05i $, $f_a = 0.25$, and  $U = 5$.
  }
\end{figure}

The effect of the relative permittivity of the active component material
on the simultaneous exhibition of dissipation and amplification    is taken up through
Fig.~\ref{eps_v_ea}. Therein, the real and imaginary parts $\eps^\perp_{Br}$  and
 $\eps^\parallel_{Br}$, as well as
the product  $\mbox{Im} \lec \eps^\perp_{Br} \ric \mbox{Im} \lec \eps^\parallel_{Br} \ric $, are plotted against the real and (negative) imaginary parts of $\eps_a$,
the product being negative   if and only if  $\mbox{Im} \lec \eps^\perp_{Br} \ric$ and  $\mbox{Im} \lec \eps^\parallel_{Br} \ric$ have opposite signs.
For calculating these results, we fixed $\eps_b = 3 + 0.05i $, $f_a = 0.25$, and  $U = 5$.
The $\eps_a$-regime for which the HCM exhibits simultaneously both amplification and attenuation is characterized by relatively small values of  $\mbox{Re} \lec \eps_a \ric $
 and relatively large values of $- \mbox{Im} \lec \eps_a \ric $. Indeed, as $ - \mbox{Im} \lec \eps_a \ric $ approaches $0$, the HCM is either exclusively dissipative or exclusively active,
 this  state being attained at larger values of $ - \mbox{Im} \lec \eps_a \ric $
as $ \mbox{Re} \lec \eps_a \ric $ approaches $2$.

Next let us turn to a more sophisticated  homogenization formalism: the SPFT \c{MAEH}. Unlike the Bruggeman  formalism, the SPFT can accommodate a comprehensive description of the distributional statistics of the component materials, via the characteristic functions
\begin{equation}
\Phi_{ \ell}(\#r) = \left\{ \begin{array}{ll} 1, & \qquad \#r \in
V_{\, \ell},\\ & \qquad \qquad \qquad  \qquad \ell\in\lec{a,b}\ric.  \\
 0, & \qquad \#r \not\in V_{\, \ell}, \end{array} \right.
\end{equation}
Herein,
 the regions occupied by the component materials `a' and `b' are identified by $V_a$ and $V_b$, respectively.
 The   ensemble average  $\langle \, \Phi_{\ell}(\#r) \, \rangle_e$ equals the volume fraction $f_\ell$,  $\ell\in\lec{a,b}\ric$.
 In the usual implementation of the  bilocally approximated SPFT
 \c{MAEH}, the distributional statistics of the component materials are characterized by the
  second moment
  \begin{equation} \l{step_cov}
\langle\Phi_\ell (\#r)\Phi_\ell (\#r')\rangle_e = \left\{
\begin{array}{lcr}
f_\ell, & & \left| \=U^{-1} \. \le \#r - \#r' \ri \right| \leq L,\\
\vspace{-6pt} && \\
f^2_\ell, & & \left| \=U^{-1} \. \le \#r - \#r' \ri \right| >
L,
\end{array}
\right.
\end{equation}
 for $ \ell\in\lec{a,b}\ric$, with
  $L$ as
  the  correlation length.

\begin{figure}[!ht]
\centering
\includegraphics[width=7.0cm]{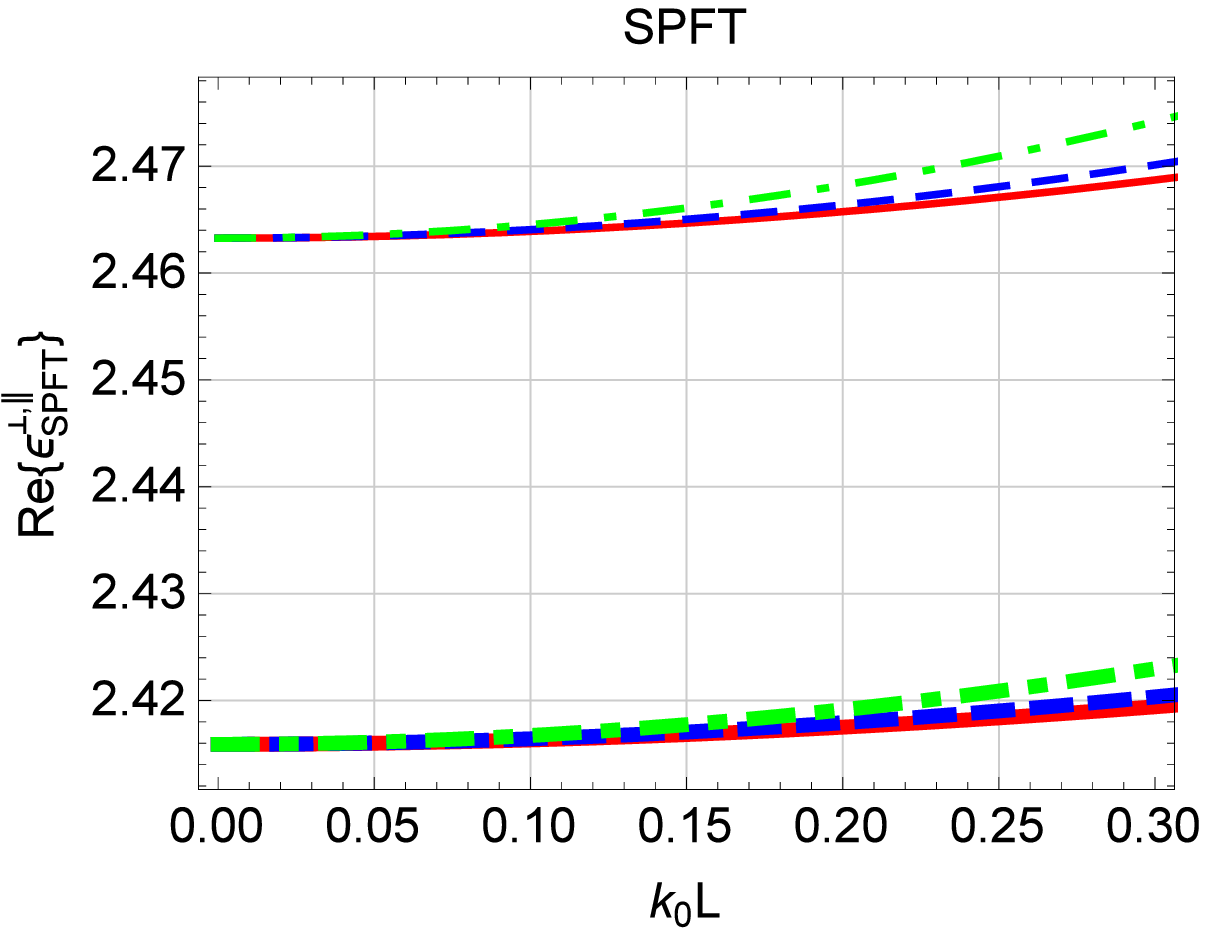}\\
 \includegraphics[width=7.0cm]{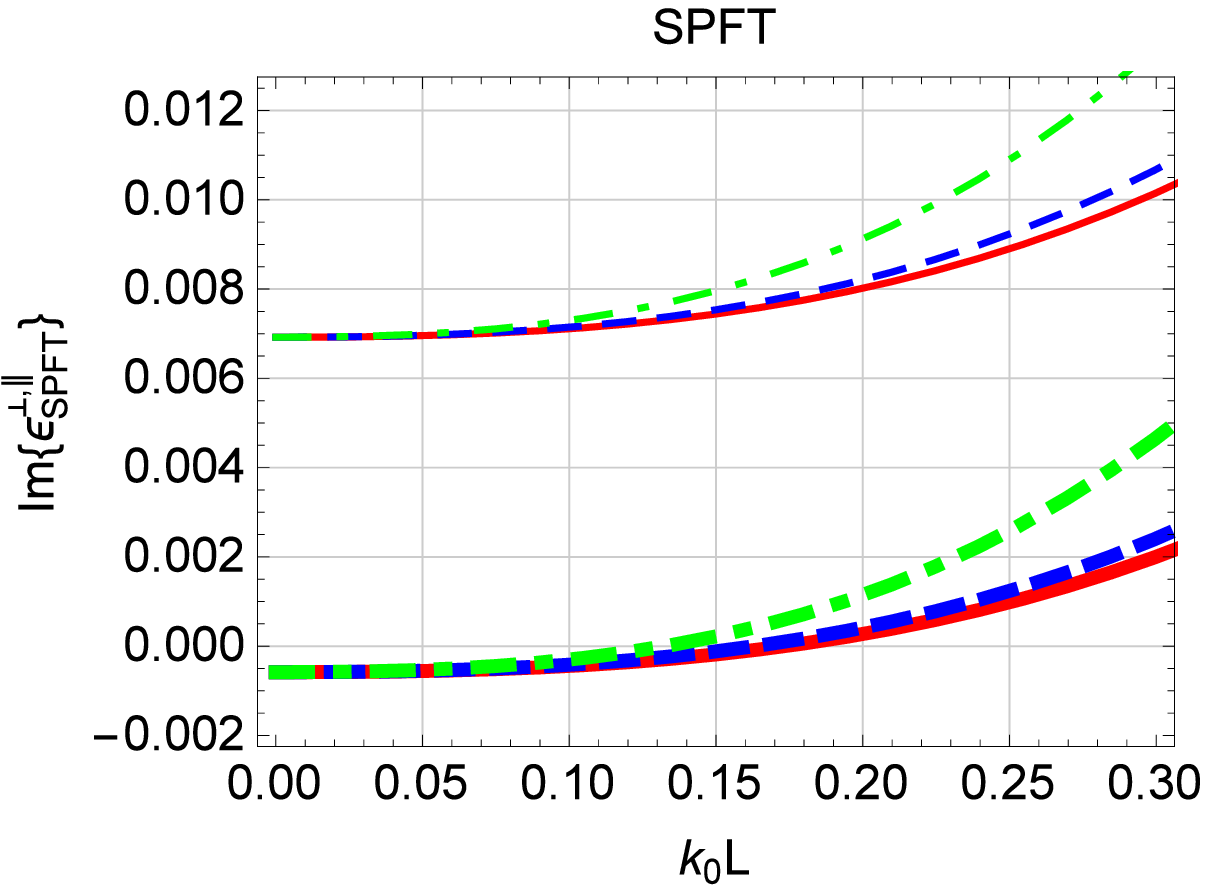}
 \caption{\label{eps_v_L} (Color
  online)
Real and  imaginary parts of  $\eps^\perp_{SPFT}$ (thick curves) and
 $\eps^\parallel_{SPFT}$ (thin curves)
 plotted against the normalized correlation length  $\ko L$, when $\eps_a = 2 - 0.03 i $, $\eps_b = 3 + 0.05i $,
 $f_a = 0.535$, and   $U = 5$.
  The  size parameter $\rho =  0$ (red, solid curves),  $0.5L$ (blue, dashed curves),
  and $L$ (green, broken dashed curves).
 }
\end{figure}

The real and imaginary parts of the HCM's relative-permittivity scalars $\eps^\perp_{SPFT}$  and
 $\eps^\parallel_{SPFT}$, as estimated using the bilocally approximated SPFT,
 are plotted against the normalized correlation length  $\ko L$ in Fig.~\ref{eps_v_L}, where $\ko = \omega \sqrt{\epso \muo}$.
 For these calculations, we set
 $\eps_a = 2 - 0.03 i $, $\eps_b = 3 + 0.05i $, $f_a = 0.535$, and   $U = 5$; also, we used an
 extended version of the SPFT \cite{eSPFT}
which explicitly accommodates the  particle--size parameter $\rho$.
Results are presented in Fig.~\ref{eps_v_L} for $\rho \in \lec 0, 0.5L, L \ric$.
In this figure, the real and imaginary parts of both $ \eps^\parallel_{SPFT} $
and $ \eps^\parallel_{SPFT} $
increase uniformly as $\ko L$ increases, for all values of  $\rho/L$ considered.
Also,
 $\mbox{Im} \lec \eps^\parallel_{SPFT} \ric>0$    for all values of $\ko L$ and $\rho/L$. However,
$\mbox{Im} \lec \eps^\perp_{SPFT} \ric<0$   for low values of $\ko L$
 while $\mbox{Im} \lec \eps^\perp_{SPFT} \ric>0$  for high values of $\ko L$. Furthermore, the transition from negative $\mbox{Im} \lec \eps^\perp_{SPFT} \ric$ to positive   $\mbox{Im} \lec \eps^\perp_{SPFT} \ric$ occurs at lower values of $\ko L$ when the size parameter $\rho$ is larger.
Thus, the HCM exhibits simultaneously both amplification and dissipation provided that both $L$ and $\rho$ are  sufficiently small.

\begin{figure}[!ht]
\centering
\includegraphics[width=7.5cm]{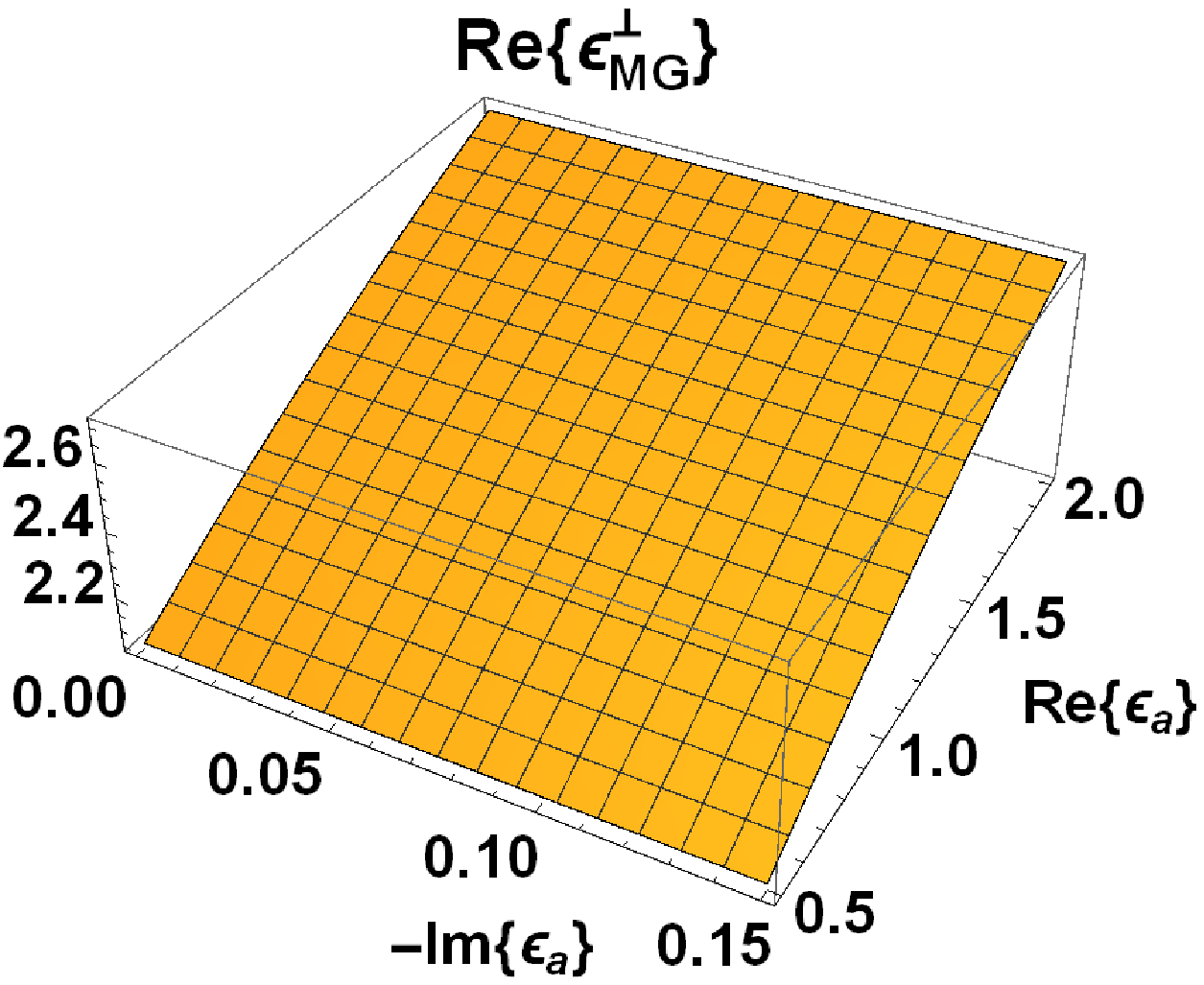}
\includegraphics[width=7.5cm]{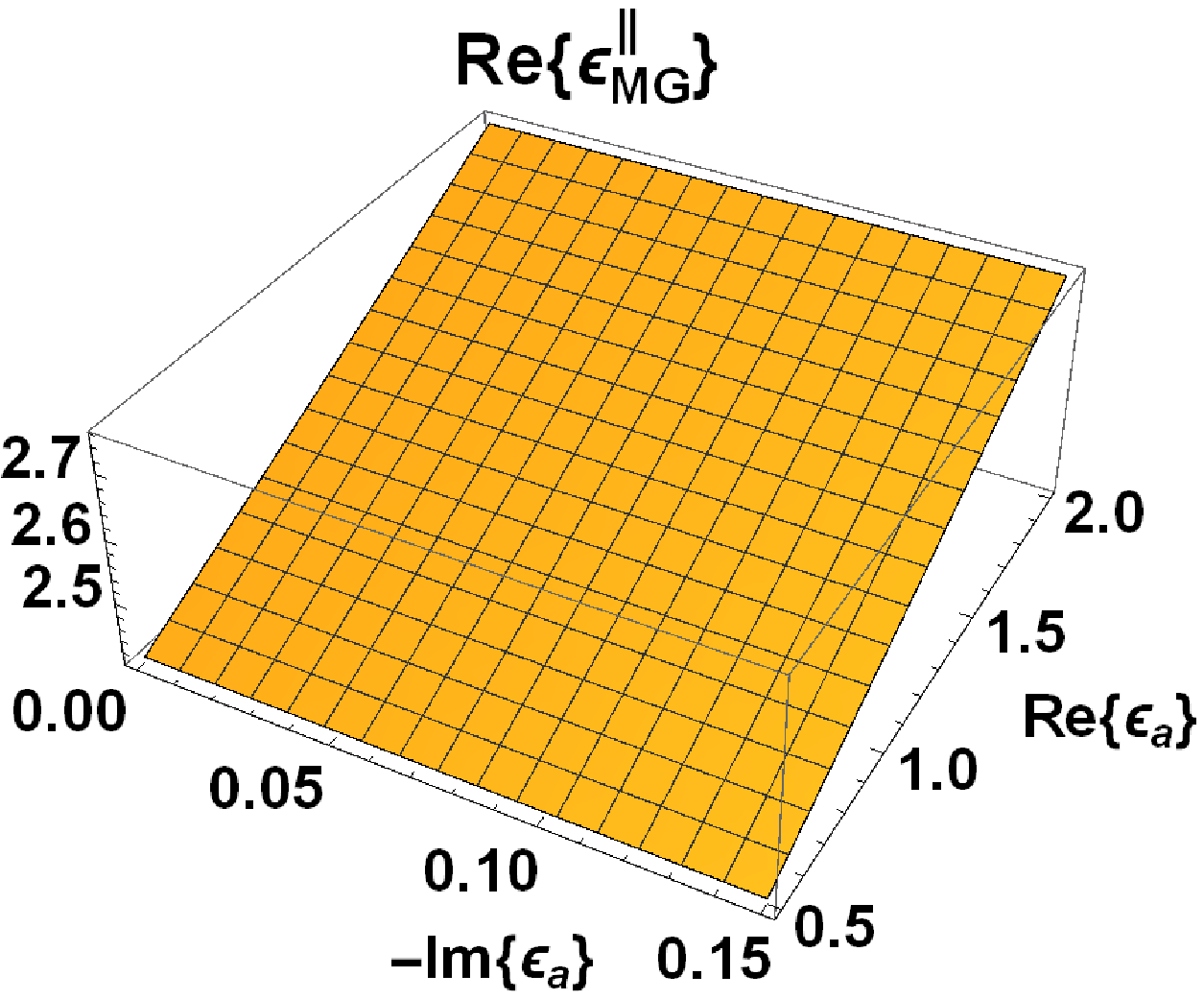} \\
\includegraphics[width=7.5cm]{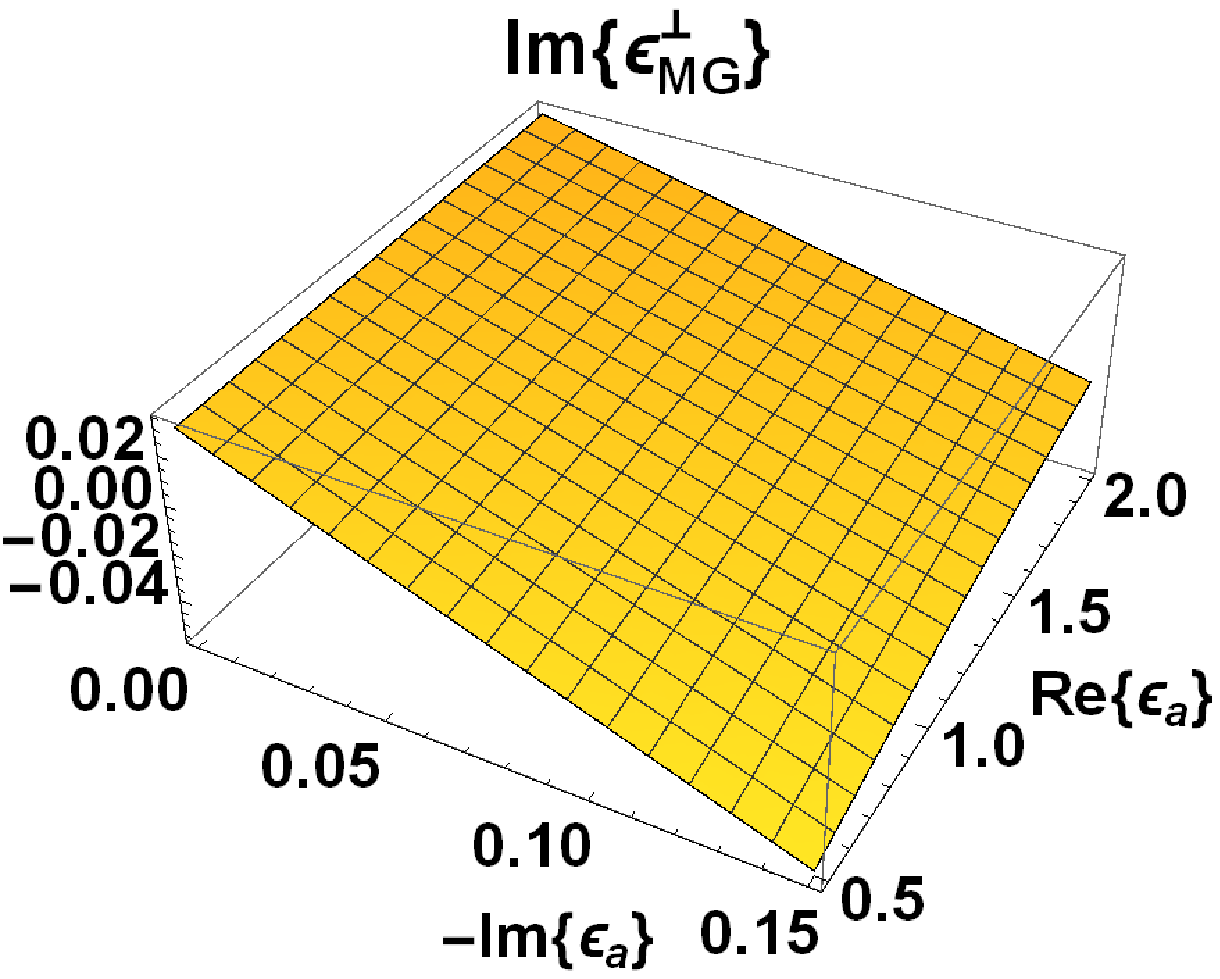}
\includegraphics[width=7.5cm]{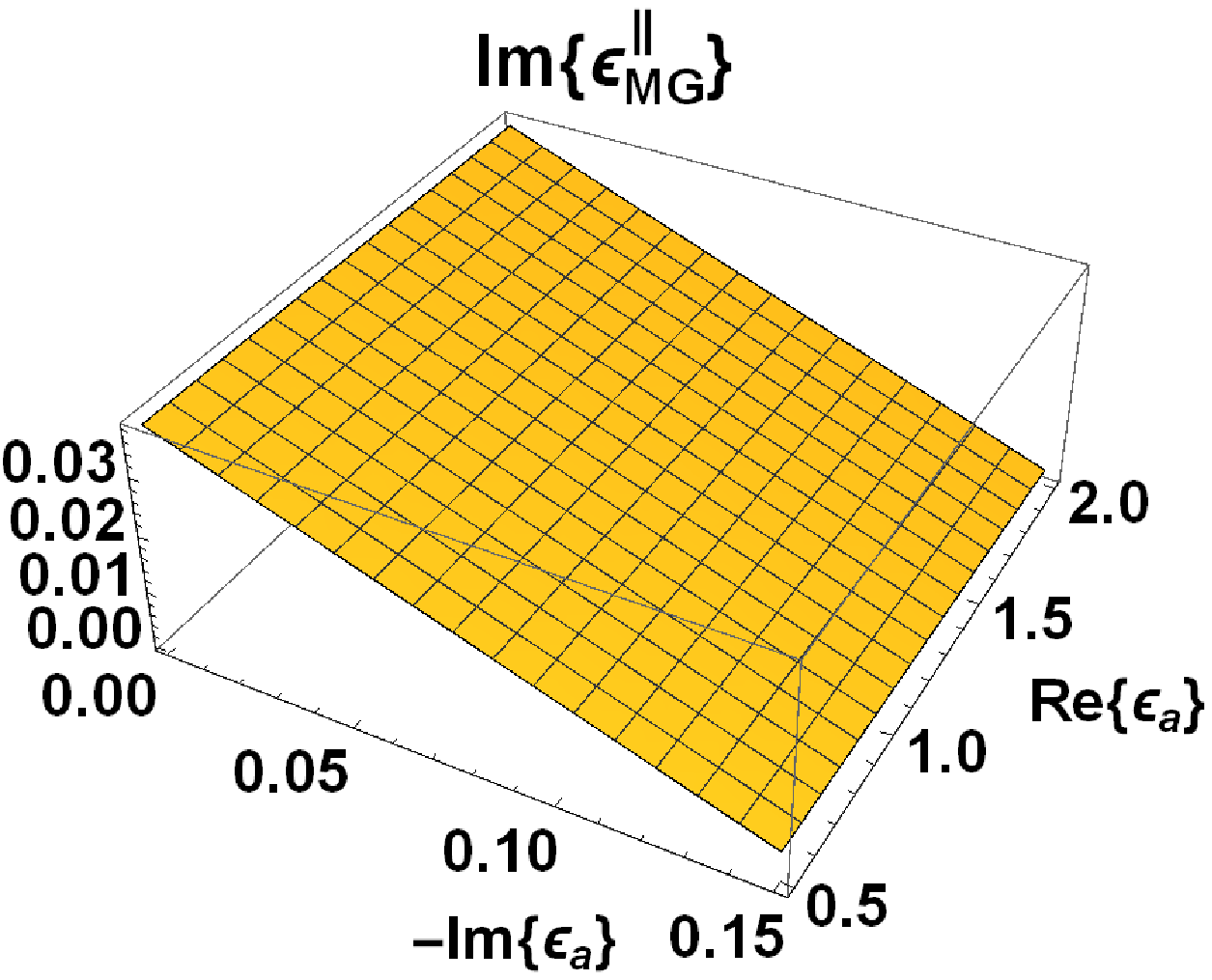}\\
\includegraphics[width=7.5cm]{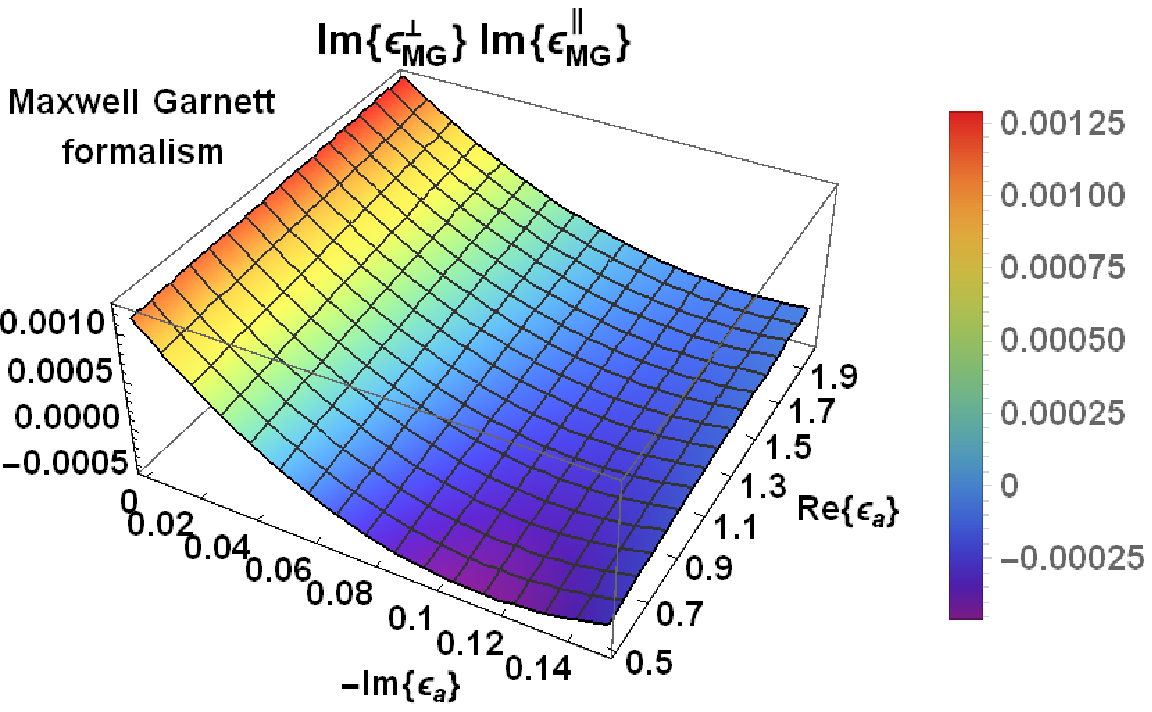}
 \caption{\label{eps_v_ea_MG} (Color
  online)
 Real and imaginary parts $\eps^\perp_{MG}$  and
 $\eps^\parallel_{MG}$, as well as
 the product  $\mbox{Im} \lec \eps^\perp_{MG} \ric \mbox{Im} \lec \eps^\parallel_{MG} \ric $,
 plotted against
  the real and (negative) imaginary parts of the relative permittivity $\eps_a$, when
 $\eps_b = 3 + 0.05i $, $f_a = 0.25$, and  $U = 5$.
Particles of component material `a' are taken to be dispersed randomly in the component material `b',
and the shape parameter $U$ applies only to component material `a'.
}
\end{figure}

Lastly, we present estimates of the HCM's relative permittivity dyadic provided by the Maxwell Garnett formalism \c{WLM_MOTL}, whose provenance is quite different from that of either the Bruggeman formalism or the SPFT. We consider
that particles of component material `a' are dispersed randomly in the component material `b'.
The Maxwell Garnett formalism is restricted to dilute composite materials ($f_a \lessapprox 0.3$).  The
 real and imaginary parts $\eps^\perp_{MG}$  and
 $\eps^\parallel_{MG}$, as well as
 the product
  $\mbox{Im} \lec \eps^\perp_{MG} \ric \mbox{Im} \lec \eps^\parallel_{MG} \ric $, are plotted against
  the real and (negative) imaginary parts of $\eps_a$ in Fig.~\ref{eps_v_ea_MG}.
 The component materials are as specified for  Fig.~\ref{eps_v_ea}, with one exception:  the shape parameter $U$ applies only to component material `a' since particle shape is irrelevant to the host material in the Maxwell Garnett formalism.
The extent of the $\eps_a$-regime for which $\mbox{Im} \lec \eps^\perp_{MG} \ric \mbox{Im} \lec \eps^\parallel_{MG} \ric < 0$ in
Fig.~\ref{eps_v_ea_MG}  is slightly smaller that
the analogous $\eps_a$-regime  in Fig.~\ref{eps_v_ea}. But, other than this relatively minor difference, the plots
 of $\mbox{Im} \lec \eps^\perp_{MG} \ric \mbox{Im} \lec \eps^\parallel_{MG} \ric $ in
Fig.~\ref{eps_v_ea_MG} and of
$\mbox{Im} \lec \eps^\perp_{Br} \ric \mbox{Im} \lec \eps^\parallel_{Br} \ric $ in Fig.~\ref{eps_v_ea} are very similar.

The estimates of $ \=\eps_{\,HCM}$ provided by the Bruggeman and Maxwell Garnett formalisms---as
represented in Figs.~\ref{eps_v_fa},
\ref{eps_v_Ua},
\ref{eps_v_ea}, and
\ref{eps_v_ea_MG}---are in close agreement over the volume-fraction range appropriate to the Maxwell Garnett formalism (i.e., $0 < f_a \lessapprox 0.3 $). The  estimates provided by the two formalisms are identical in the limit $f_a \to 0$ and very small differences emerge as $f_a$ increases. Indeed, the
corresponding
 plots of the
real and imaginary parts of the components of $ \=\eps_{\,Br}$ and  $ \=\eps_{\,MG}$ in Figs.~\ref{eps_v_ea} and \ref{eps_v_ea_MG}, respectively,
are almost indistinguishable to the naked eye.
 These estimates are also in close agreement
 with the corresponding estimates   provided by the bilocally approximated SPFT, as represented in Fig.~\ref{eps_v_L}.
 The imaginary parts of the components of  $ \=\eps_{\,HCM}$ estimated by the SPFT deviate slightly
 from their corresponding Bruggeman and Maxwell Garnett counterparts, with the deviation increasing in magnitude as the correlation length $L$
  and the size parameter $\rho$ increase. This deviation reflects the fact that the SPFT formalism accommodates coherent scattering losses via $L$ and $\rho$ \c{MAEH}, whereas the Bruggeman and Maxwell Garnett formalisms do not.

 Rigorous theoretical bases have been firmly established for
  each of the  three homogenization formalisms employed here \c{WLM_MOTL,Kong,Z94,David,JW,MAEH}.
 Techniques based on such  homogenization formalisms have the advantages over full-wave numerical techniques, based on  the finite-element method or the finite-difference
 time-domain method \c{FEM_Brosseau,Monte_Carlo_FEM_JAP,FD_bianisotropic}, for example, that
  they provide estimates of
the constitutive parameters which are independent of the shape and size of the bulk
material involved, and these estimates apply for all possible incident fields with sources not located in the bulk material. However, it is important to bear in mind that the predictions of constitutive parameters provided by any
homogenization formalism are \emph{estimates}. The ultimate checks on such estimates  can only be provided by careful
experimental studies.

\section{SIMULTANEOUS ATTENUATION AND AMPLIFICATION Exemplified}

Power flow associated  with electromagnetic  propagation  is represented by the time-averaged Poynting vector $ {\#P}   = \le 1/2 \ri \mbox{Re} \lec \#E   \times \#H^*  \ric$, with  the operator $\mbox{Re} \lec \cdot \ric$ delivering the real part.
Since $ \#\nabla \.{\#P}    =- Q$ in a region devoid of externally impressed sources \cite{Zangwill,Kong-Book}, amplification and dissipation should be discernible through plane-wave propagation.

Suppose that the direction of propagation is parallel to the unit vector $\hat{\#a}$. An
ordinary plane wave  will propagate in the chosen HCM with wavenumber $k_{or}=\ko \sqrt{\eps^\perp_{HCM}}$,
regardless of the angle $\theta=\cos^{-1}\le\hat{\#a}\.\hat{\#u}\ri$. An
extraordinary plane wave  will propagate in the chosen HCM with wavenumber
\begin{equation}
k_{ex}=\ko \sqrt{\frac
{\eps^\perp_{HCM} \eps^\parallel_{HCM}}
{\eps^\perp_{HCM}\sin^2\theta+ \eps^\parallel_{HCM}\cos^2\theta}}
\end{equation}
that does depend on $\theta$.

Consider two examples: (i) When {$f_a=0.6$ in Fig.~\ref{eps_v_fa}, $\eps^\perp_{Br} = 2.353 - 0.006 i$ and
$\eps^\parallel_{Br} = 2.398 + 0.002 i$, which yield $k_{or} = \le 1.534 - 0.002 i \ri \ko$
with $\mbox{Re} \lec k_{ex} \ric > 0$ for all $\theta$ and $\mbox{Im} \lec k_{ex} \ric > 0$
for $ 60.5^\circ < \theta < 119.5^\circ$. Hence, the ordinary plane wave is amplified for all propagation directions while the extraordinary plane wave is amplified for $\theta \in \le 0^\circ, 60.5^\circ \ri \cup \le 119.5^\circ, 180^\circ \ri $ but attenuated for $\theta \in \le  60.5^\circ, 119.5^\circ \ri$.
(ii) When $U=0.25$ in Fig.~\ref{eps_v_Ua}, $\eps^\perp_{Br} = 2.456 + 0.006 i$ and
$\eps^\parallel_{Br} = 2.383 - 0.006 i$, which yield $k_{or} = \le 1.567 + 0.002 i \ri \ko$
with $\mbox{Re} \lec k_{ex} \ric > 0$ for all $\theta$ and $\mbox{Im} \lec k_{ex} \ric < 0$
for $ 44.1^\circ < \theta < 135.9^\circ$. Hence, the ordinary plane wave is attenuated for all propagation directions while the extraordinary plane wave is amplified for $\theta \in \le  44.1^\circ, 135.9^\circ \ri$  but attenuated for  $\theta \in \le 0^\circ, 44.1^\circ \ri \cup \le 135.9^\circ, 180^\circ \ri $.\\

\section{Closing remarks}

In conclusion, according to the estimates afforded by three different well--established homogenization formalisms,
an HCM that  exhibits simultaneously   attenuation and amplification of electromagnetic fields at a specific frequency may be realized quite simply as a random mixture of electrically small spheroids  of  two different materials. Both component materials are isotropic dielectric materials, one of which is dissipative while the other is active.  The realization of such an HCM depends upon the volume fraction,  spatial distribution, particle shape and size, and the relative permittivities of the component materials. Dynamic control of the active component material, for example via stimulated Raman scattering,  affords dynamical control of the HCM.
Thus, a new class of  metamaterials is proposed which are neither wholly dissipative nor wholly active. Although we have illustrated the concept with uniaxial dielectric HCMs, particulate composite materials with more complicated linear and/or nonlinear constitutive properties and displaying both dissipation and amplification at the same frequency can be designed \cite{EAB}. Finally, even periodically laminated composite materials
\c{BW}
offer similar promise, but particulate composite materials
 may be more readily fabricated than their laminar counterparts.

\vspace{2mm}

\noindent {\bf Acknowledgments:} TGM acknowledges the support of EPSRC grant EP/M018075/1. AL thanks the Charles Godfrey
Binder Endowment at Penn State for partial financial support of his
research activities.

\end{document}